\documentstyle[preprint,psfig]{aastex}

\def\lesssim{\mathrel{\hbox{\rlap{\hbox{\lower4pt\hbox{$\sim$}}}\hbox{$<$}}}}
\def\gtrsim{\mathrel{\hbox{\rlap{\hbox{\lower4pt\hbox{$\sim$}}}\hbox{$>$}}}}

\begin{document}

\title{Global MHD Simulation of the Inner Accretion Disk in a 
Pseudo-Newtonian Potential}

\author{John F. Hawley}
\affil{Department of Astronomy, University of Virginia,
Charlottesville VA 22903}

\author{Julian H. Krolik}
\affil{Physics and Astronomy Department, Johns Hopkins University,
    Baltimore, MD 21218}

\shorttitle{Global MHD Simulation}

\begin{abstract}

We present a detailed three dimensional magnetohydrodynamic (MHD)
simulation describing the inner region of a disk accreting onto a black
hole.   To avoid the technical complications of general relativity, the
dynamics are treated in Newtonian fashion using the pseudo-Newtonian
Pacz\'ynski-Wiita potential.  The disk evolves due to angular momentum
transport that is produced naturally from MHD turbulence generated by
the magnetorotational instability.  We find that the resulting stress
is continuous across the marginally stable orbit, in contradiction
with the widely-held assumption that the stress should go to
zero there.  As a consequence,
the specific angular momentum of the matter accreted into the hole is
smaller than the specific angular momentum at the marginally stable
orbit.  The disk exhibits large fluctuations in almost every quantity,
both spatially and temporally.  In particular, the ratio of stress to
pressure (the local analog of the Shakura-Sunyaev $\alpha$ parameter)
exhibits both systematic gradients and large fluctuations; from $\sim
10^{-2}$ in the disk midplane at large radius, it rises to $\sim 10$
both at a few gas density scale heights above the plane at large radius,
and near the midplane well inside the plunging region.   Driven in part
by large-amplitude waves excited near the marginally stable orbit, both
the mass accretion rate and the integrated stress exhibit large
fluctuations whose Fourier power spectra are smooth ``red" power-laws
stretching over several orders of magnitude in timescale.

\end{abstract}

\keywords{accretion, accretion disks, MHD, black holes}

\section{Introduction}

For many years, the physical nature of angular momentum transport
in accretion disks has been one of the great outstanding
questions of high-energy astrophysics.  As a result of nearly a decade
of effort, it is now becoming increasingly clear that the mechanism is
turbulent magnetic stress in which the energy for the magnetic field is
largely drawn from the free energy of orbital shear (as reviewed by
Balbus \& Hawley 1998).  Theoretical grounds for this conclusion were
first found in the rediscovery of the magneto-rotational instability
(MRI) by Balbus \& Hawley (1991).  Increasingly detailed simulations of
local ``shearing boxes'' have bolstered our confidence that this is
indeed the process found in nature (e.g., Hawley, Gammie \& Balbus
1995, 1996; Brandenburg et al. 1995; Stone et al. 1996; 
Miller \& Stone 2000).

What has been missing until recently is studies of the radial structure
of disks accreting under the influence of magnetohydrodynamic (MHD)
turbulent stresses.  Hitherto, only a few such three-dimensional global
simulations have been presented (Armitage 1998; Matsumoto 1999; Hawley
2000; Machida, Hayashi, \& Matsumoto 2000).  
In global simulations it is difficult to obtain adequate
resolution over the wide range of length scales involved, namely the
MHD turbulent length scales within the disk, the disk scale height $H$,
and the radial distance from the central object $R$.  These length
scales are generally quite disparate, but near the inner boundary of an
accretion disk they should become comparable.  In this paper we present
a global simulation of an accretion disk in which we focus on this
inner region where adequate numerical resolution can be obtained.

With the results of this new simulation, we can begin to answer a
number of questions of central interest to the mechanics of accretion.
For example, the most common view of the inner portions of black hole
accretion disks is based on the work of Novikov \& Thorne (1973) (e.g.
Abramowicz \& Zurek 1981; Muchotrzeb \& Paczy\'nski 1982; Matsumoto et
al. 1984).  In this picture, the disk is (nearly) time-steady and
axisymmetric.  Page \& Thorne (1974) argued on heuristic grounds that
the $R$-$\phi$ component of the stress (the one responsible for angular
momentum transport) should go to zero at the marginally stable orbit;
later work by Abramowicz and Kato (1989) showed that if the stress is a
(constant) small fraction of the local pressure, it automatically
approaches zero near that location.

However, because the MHD turbulence is the result of a general, local,
rapidly-growing instability, assumptions such as stationarity and
axisymmetry are likely inappropriate.  One might also question whether
there is any reason for the stress to diminish at the disk's inner
edge, given that the growth rate of the MRI does not appreciably
diminish near the marginally stable orbit (cf. Krolik 1999; Gammie 1999;
see also the footnote in Page \& Thorne 1974 in which they speculate
that their heuristic arguments might fail if strong magnetic fields were
present).  Further, because magnetic
stresses do not automatically maintain a fixed ratio to the pressure,
conclusions based upon $\alpha$ parameterizations (e.g., Abramowicz \&
Kato 1989) may not directly apply.  This issue is important because
continued stress at the marginally stable orbit could alter both the
energy and angular momentum with which matter arrives at the black
hole; that is, this boundary condition determines the efficiency of
accretion and the rate at which the black hole is spun up.

Because the greatest luminosity should be released near the inner edge
of the accretion disk, the detailed behavior and structure of the
region near the marginally stable orbit is crucial to
understanding the observations of black hole systems.  For example, the
luminosity of essentially all accreting black holes exhibits sizable
fluctuations with a very broad-band distribution of fluctuation power
with timescale (e.g., as discussed by Sunyaev \& Revnivtsev 2000); can we
identify the specific dynamics that drive these variations?
Occasionally, the lightcurve exhibits
quasi-periodic oscillations (QPOs); can we either identify the
mechanism responsible for these, or test some of those mechanisms that
have been proposed?

In the following section we will present a technical description of the
three-dimensional global simulation we report.  We will then set out
our results in \S 3, with their qualitative implications outlined in \S
4.  In \S 5 we will discuss the degree to which limitations of our
simulation hamper direct application of our results, and suggest in
which directions the greatest improvement might soon be made.
We summarize our conclusions in \S 6.

\section{Numerical Method}

\subsection{Equations}

An investigation of the inner edge of a disk orbiting a black hole
would be best carried out with a fully general relativistic
simulation.  However, we have not yet written a three dimensional
general relativistic MHD code.  In the present study we instead make
use of the existing three dimensional Newtonian MHD code (Hawley 2000).
The code evolves the equations of ideal MHD, i.e.,
\begin{equation}\label{mass}
{\partial\rho\over \partial t} + \nabla\cdot (\rho {\bf v}) =  0
\end{equation}
\begin{equation}\label{mom}
\rho {\partial{\bf v} \over \partial t}
+ (\rho {\bf v}\cdot\nabla){\bf v} = -\nabla\left(
P + {\mathcal Q} +{B^2\over 8 \pi} \right)-\rho \nabla \Phi +
\left( {{\bf B}\over 4\pi}\cdot \nabla\right){\bf B}
\end{equation}
\begin{equation}\label{ene}
{\partial\rho\epsilon\over \partial t} + \nabla\cdot (\rho\epsilon
{\bf v}) = -(P+{\mathcal Q}) \nabla \cdot {\bf v}
\end{equation}
\begin{equation} \label{ind}
{\partial{\bf B}\over \partial t} =
\nabla\times\left( {\bf v} \times {\bf B} \right)
\end{equation}
where $\rho$ is the mass density, $\epsilon$ is the specific internal
energy, ${\bf v}$ is the fluid velocity, $P$ is the pressure, $\Phi$ is
the gravitational potential, ${\bf B}$ is the magnetic field vector,
and ${\mathcal Q}$ is an explicit artificial viscosity of
the form described by Stone \& Norman (1992a).
The global disk code is written in cylindrical coordinates, $(R,\phi,z)$.  
To model important effects associated with the relativistic Schwarzschild
metric, we employ the pseudo-Newtonian potential of Paczy\'nski \& Wiita
(1980). The pseudo-Newtonian potential has the form
\begin{equation}\label{pwp}
\Phi = - {G M \over r-r_g}
\end{equation}
where $r$ is spherical radius,
and $r_g \equiv 2GM/c^2$ is the ``gravitational radius,''
akin to the black hole horizon.   For this potential,
the Keplerian specific angular momentum (i.e., that
corresponding to a circular orbit) is
\begin{equation}\label{pwl}
\ell_{kep} = (GMr)^{1/2} {r \over r-r_g} ,
\end{equation}
and the angular frequency $\Omega = \ell/R^2$.  The innermost marginally
stable circular orbit is located at $r_{ms}=3r_g$. 

We use an adiabatic equation of state, $P=\rho\epsilon(\Gamma
-1) = K\rho^\Gamma$, with $\Gamma = 5/3$, and ignore radiation
transport and losses.  Since there is no explicit resistivity or
physical viscosity, the gas can heat only through adiabatic
compression or by the action of the artificial viscosity
which acts in shocks.
Equations (\ref{mass})-(\ref{ind}) are solved using time-explicit
Eulerian finite differencing.  The numerical algorithm is that employed
by the ZEUS code for hydrodynamics (Stone \& Norman 1992a) and MHD
(Stone \& Norman 1992b; Hawley \& Stone 1995).  We adopt units where
$GM=1$ and $r_g = 1$ (so that $c = \sqrt{2}$).

In this paper we repeat---with significantly better
resolution---simulation GT4 from Hawley (2000), whose initial state was
a thick torus with an angular velocity distribution $\Omega \propto
R^{-1.68}$, slightly steeper than Keplerian.  The angular momentum
within the torus is equal to the Keplerian value at $R=10$, which
determines the location of the pressure maximum in the torus.  In
somewhat arbitrary units, the pressure at this point is equal to 0.036
and the density $\rho_{max}=34$.  The initial magnetic field is
constructed by setting the toroidal component of the vector potential
$A_\phi (R,z) = \rho (R,z) -
\rho_{min}$, for all $\rho$ greater than a minimum value, $\rho_{min} =
0.1$.  Poloidal field components are then constructed from ${\bf B} =
\nabla \times {\bf A}$.  This procedure produces large-scale poloidal
field lines that follow the equal-density contours inside the torus.
The magnetic field is initially zero everywhere $\rho \leq \rho_{min}$.
As discussed in
Hawley (2000), this configuration is motivated by the desire to have
well-resolved unstable modes of the poloidal field instability, and to
have the initial field fully contained within the torus.  The magnetic
field level is renormalized so that the total magnetic energy in the
torus is 0.01 the total integrated gas pressure.  This corresponds to an
overall value of $\beta=P_{\rm gas}/P_{\rm mag} = 100$.

  The aim of this initial condition is not so much to
follow the evolution of this particular torus, as it is to construct a
disk that is nearly Keplerian with an inner boundary that is initially
close to, but still outside, the marginally stable orbit.  To that end,
the torus is perturbed with low-amplitude random adiabatic pressure
fluctuations.  These seed the magneto-rotational instability whose
rapid evolution leads to the desired state.  Further disk evolution
results in inflow through $r_{ms}$;
it is this resulting accretion flow that will be studied in detail.

The computational grid has $128\times 128 \times 128$ zones.  This is the
same number of zones as used in GT4 of Hawley (2000), but here,
to increase the resolution within the disk itself and in the inflow
region,  we locate more of the zones near the marginally stable orbit
and around the equator.  The center of the coordinate system is
excised, i.e., the radial coordinate begins at a nonzero
$R_{\rm min} =1.5$; this choice avoids both the coordinate singularity
associated with the axis and the singularity at $r_g$.  The outer
radius is set at $R=31.5$.  The $z$ coordinate is centered on the
equatorial plane, and runs from -10 to $+10$.  The angle $\phi$ is
periodic and covers the full $2\pi$.  In the radial direction, there
are 30 equally spaced zones in $R$ from 1.5 to 4; the radial zones are
then graded logarithmically from 4 to the outer boundary at 31.5.  In
the vertical direction, half of the $z$ zones are equally spaced around
the equator from $-2$ to 2, then graded from those locations out to the
vertical boundaries at $z = \pm 10$.  The grid cells are equally spaced
in $\phi$.  The simulation GT4 from Hawley (2000) used equally spaced
zones in $R$ and $z$, and the outer radial boundary was set at 21.5.
We have also run this initial torus at lower resolution for comparison,
using $64^3$ grid zones.

The boundary conditions on the grid are simple zero-gradient outflow
conditions; no flow into the computational domain is permitted.  The
magnetic field boundary condition is set by requiring the transverse
components of the field to be zero outside the computational domain,
while the perpendicular component satisfies the divergence-free
constraint.  Although this produces a nonzero stress at the boundary,
it works well in preventing an artificial buildup of field at the
boundary.  Most of the time, for $|z| \leq 1$ along the inner radial
boundary and $R \leq 2$ in the equatorial plane, the poloidal velocity
is greater than the magnetosonic speed, so that the influence of the
boundary condition on the flow upstream is limited.

Timescales of interest in the simulation are set by the circular
orbital period $P_{orb} = 2\pi \Omega^{-1} = 2\pi r^{3/2} (r-r_g)/r$ at
significant points within the disk.  In units where $GM=r_g=1$,
$P_{orb}$ at the pressure maximum ($R=10$) is $179$, and at the
marginally stable orbit, where the orbital frequency $\Omega =
1/(2\sqrt 3)$, the orbital period is 21.8.  The local accretion
timescale is set by the effective stress, namely $t_{acc} = R/\langle
v_R \rangle \sim \ell/W_{R\phi} ( \gg P_{orb})$, where $W_{R\phi}$ is
the $R$--$\phi$ component of the stress, $W_{R\phi} = \langle \delta
v_R \, \delta v_\phi - {v_{AR} v_{A\phi} \rangle}$, and $v_A$ is an
Alfv\'en velocity.  The goal is to evolve long enough to observe a
number of accretion times near the marginally stable orbit.  Since the
code is time-explicit and therefore Courant-limited, this is difficult
to achieve over large radial extents.  Here we ran the simulation out
to time $t=1500$, which is 69 orbits at $r_{ms}$.

\subsection{Diagnostics}

The quantity of data associated with a three dimensional simulation is
daunting, and this makes analyzing the outcome a challenge.  Diagnostics
can be computed during the simulation and in post-processing.  The best 
procedures for this are still being developed.  Here we briefly describe
some of the diagnostic procedures used in the present simulation.

The time-evolution of the disk can be visualized through animation
sequences.  Animation frames of the $\phi = 0$ slice and the equatorial
($z=0$) slice are dumped every 4 units in time.  Full 3D density
information is saved every 10 units of time, and these are used to make
an animation of a three dimensional volumetric rendering.

The time-evolution can also be studied in a manageable way through
space-time $(R,t)$ studies of azimuthally- and vertically-integrated quantities.
Examples include the averaged mass density 
\begin{equation}\label{sigma}
\langle \rho\rangle = {\int \rho R d\phi dz \over \int R d\phi dz},
\end{equation}
the average angular momentum,
\begin{equation}\label{angm}
\langle \rho \ell \rangle  = {\int \rho \ell R d\phi dz \over \int R
d\phi dz},
\end{equation}
and average specific angular momentum,
\begin{equation}\label{el}
\langle \ell \rangle= {\langle \rho \ell \rangle \over \langle\rho\rangle }.
\end{equation} 
The net radial mass flux is
\begin{equation}\label{mdot}
\langle \dot M\rangle = \langle -\rho v_R \rangle = \int -\rho v_R R d\phi dz,
\end{equation}
here defined to be positive for net accretion (inflow).
Similarly, the angular momentum flux,
\begin{equation}\label{lflux}
\langle \rho v_R \ell \rangle = \int \rho v_R \ell R d\phi dz,
\end{equation}
can be computed.

Radial angular momentum transport in the disk is due to
the $R$--$\phi$ component of the stress,
\begin{equation}\label{stress}
T^{R\phi} = \rho \delta v_R \, \delta v_\phi - {B_R B_\phi \over 4\pi},
\end{equation}
where $\delta v_R$ and $\delta v_\phi$ are the {\it turbulent} portions
of the velocity, that is, the departures from the mean flow.  The
kinematic portion is the Reynolds stress and the magnetic portion is
the Maxwell stress.  In the simulation we have no way of
knowing what the mean flow is.   For example, the typical radial
velocity is much larger than the time-averaged net radial drift
velocity, and there are significant differences between the
instantaneous angular velocity and the value associated with a circular
orbit (which one might assume should be the mean orbital flow).  As a
result, there is no  unique prescription to compute the Reynolds
component of the total stress.  Hawley (2000) used a definition for the
perturbed orbital velocity $\delta v_\phi$ in terms of the difference
between the total instantaneous angular momentum flux, and the mass
flux times the average specific angular momentum.  The difference then
represents the excess or deficit angular momentum transport due to
orbital velocity fluctuations compared to the mean.  Specifically,
\begin{equation}\label{rstress}
\langle \rho \delta v_R \delta v_\phi \rangle =
\langle \rho v_R v_\phi \rangle
- \langle \rho v_R \rangle \langle \ell \rangle / R .
\end{equation}

In contrast, we can easily compute the stress due solely to magnetic
forces, the second term on the right-hand-side of equation
(\ref{stress}); we denote this quantity by $M^{R\phi}$.  The
vertically- and azimuthally-averaged Maxwell stress is
\begin{equation}\label{maxwell}
\langle M^{R\phi}\rangle = {\int (-B_R B_\phi / 4\pi) R d\phi dz
\over \int Rd\phi dz} .
\end{equation}
We do not distinguish the total magnetic field from the fluctuating field
because, as we shall show later, where the field is dynamically important,
the magnitude of the fluctuating portion is large compared to any time-averaged
mean field.

In addition to these vertically- and azimuthally-averaged quantities,
we examine full data dumps of the simulation at specific moments in
time.  From these we can examine not only the full three dimensional
structure, but also azimuthally-averaged and vertically-averaged
slices in $(R,z)$ and $(R,\phi)$.

\section{Results}

\subsection{Quasi-stationary state}

The initial torus was in hydrodynamical equilibrium, but it evolves
rapidly as the simulation proceeds.  The overall evolution is very
similar to that of GT4 in Hawley (2000; see Fig. 13).
The radial field is sheared by the differential rotation, creating
toroidal field.  Turbulence develops within the disk with the onset of
the magnetorotational instability.  The resulting Maxwell stresses
drive angular momentum transfer; the disk begins to accrete into the
central hole and expands radially outward.  Relatively rapidly the
inner portion of the disk achieves an approximate steady-state.  By $t
\simeq 700$, the mass accretion rate reaches roughly the long-time
mean; after $t \simeq 1000$, the shape of the disk's average radial
density distribution $\langle \rho\rangle (R)$ no longer changes
appreciably.  Thus, we observe the disk in a quasi-stationary state
for several dozen orbital periods at the marginally stable orbit.
In much of what follows, we will present a detailed analysis of the
state of the disk during this time frame, focusing particularly on
the end of the simulation, using that ``snapshot" as a fairly typical
sample of conditions in the quasi-stationary disk.
However, it must be borne in mind that there are always
significant fluctuations around all these mean states.

\subsection{Gas density}

A contour plot of the azimuthally-averaged gas density at the initial
and final times is presented in Figure~1.  As might be expected, the
density diminishes inside the marginally stable orbit and outside the
initial outer radial boundary of the disk, while also decreasing with
increasing altitude away from the equatorial plane.  Although the
vertical heights of the highest contour levels at the final time are
roughly independent of radius within the disk proper, the
azimuthally-averaged exponential scale-height of the pressure, $H$,
increases very nearly linearly with radius from $R \simeq 2$ out to $R
\simeq 20$; at the marginally stable orbit ($R = 3$), $H \simeq 0.4$,
while at $R = 10$, $H \simeq 1.5$.  This particular disk shape is
almost certainly a consequence of the equation of state, in which all
fluid elements retain the same specific entropy except for shock
heating.
As a result of this special choice of equation of state,
the temperature rises inward as the gas is compressed; by coincidence,
this occurs at such a rate as to keep the ratio of sound speed to
orbital speed, and therefore $H/R$, almost constant at $\simeq 0.15$.
In a real disk, one in which turbulent dissipation introduces
heat and radiation vents it, the shape might be rather different.

Density gradients in radius and altitude are determined primarily
by the overall dynamical effects of approximate hydrostatic equilibrium.
No such requirements exist for azimuthal fluctuations, whose
amplitudes characterize the turbulence.  We
quantify the azimuthal fluctuation level by defining
\begin{equation}\label{fluctuation}
{\delta \rho \over \rho}\left(R,z\right) = 
{1 \over \langle \rho\rangle_\phi}  \left\{{1 \over 2\pi} \int \, d\phi \,
             \left[\rho - \langle \rho \rangle_\phi \right]^2 \right\}^{1/2} ,
\end{equation}
where ${\langle X \rangle}_\phi$ refers to the azimuthal average at
fixed $R$ and $z$ of the quantity $X$.  Measured in this way,
$\delta\rho/\rho \simeq 0.2$ -- 0.6 over most of the problem volume,
with smaller values typically nearer the equatorial plane.  In a few
places, $\delta \rho/\rho$ rises to be greater than unity.  Thus, quite
sizable density fluctuations are generic.

    These fluctuations are primarily sheared strips whose radial extent is
$\lesssim 0.1 R$, but extend $\sim 1$~radian in azimuthal angle (Figure~2). 
This pattern (tightly wrapped spiral features) is characteristic of
fluctuations in almost all quantities.  Part of their nature is
revealed by constructing spacetime diagrams in which the surface
density at a specific radius is plotted as a function of $t$ and $\phi$.
From this visualization we are able to measure the angular speed of
these features; it is generally quite close to the local orbital frequency
$\Omega$.  We can also measure their persistence at any particular radius;
it is usually less than an orbital period.  In this sense, they can be
regarded as fluctuations that merely ``ride" with the local orbital motion,
but see further discussion in \S 3.5.

\subsection{Magnetic field}

The magnetic field distribution is rather more complicated than the
density distribution.  A poloidal projection (i.e., averaged
azimuthally) shows the magnetic pressure to have its largest value
immediately
outside the dense regions of the torus, with large variations inside
the torus (Figure~3).  In the shearing-box simulations of Miller \&
Stone (2000), the field strength likewise peaked away from the midplane,
but we find somewhat greater contrast.  Field-strength
fluctuations are also larger than
those for density: the {\it rms} fractional fluctuations in
$|\vec B|$ in
the azimuthal direction are $\simeq 1$ -- 1.5 in the disk body, falling to
$\simeq 0.25$ several gas scale-heights immediately above and below the
disk.
Another measure of the magnitude of the fluctuations is
$|\langle B_i \rangle_\phi |/\langle |B_i | \rangle_\phi$, the ratio of the
mean value of an individual field component to the mean of its absolute
value.  This quantity is typically $\sim 0.2$ for all components
wherever the gas density
is significant, indicating that the field is extremely variable as a
function of azimuth (and therefore as a function of time at fixed azimuth,
as orbital rotation carries the field around).
Only where the gas density is very low does the field
become relatively smooth and steady
(a result also qualitatively similar to the findings of Miller
\& Stone 2000).
%
Like the density, the characteristic shape of fluctuations is
sheared arcs that run for $\sim 1$~radian before losing coherence.
However, the peaks in magnetic field strength tend to lie adjacent to,
rather than coincident with, peaks in gas density.

     To illustrate some of the field structure, we show in Figure~4
poloidal slices taken at $\phi = \pi/2$.  Panel a) is a close-up of the inner,
high-resolution region; panel b) shows (almost) the entire disk.
This view, of course, suppresses the toroidal component (which is
typically larger than the poloidal component by a factor $\sim 1$ -- 10);
in addition, in order to make the figures visually comprehensible, we have
omitted small-scale field variations (smaller than 0.125 length
units in panel a), smaller than 0.5 length units in panel b).  Even
with this smoothing, it is apparent that the field is very tangled.  The
only field property exhibiting any stability is the roughly dipolar
pattern seen in the low density regions.  This is a remnant of the
initial condition.  Because it only persists in these very low density
regions, and they do not contribute much to the stress (see \S 3.6),
we do not regard this remnant artifact as dynamically significant.

As one expects from magnetic fields in shearing plasma,
$B_R$ and $B_\phi$ are correlated in the sense that the mean ratio of
the Maxwell stress to the magnetic pressure, $\langle -2 B_R B_\phi
/B^2 \rangle > 0$ and has (approximately) constant magnitude $\simeq
0.2$ -- 0.3 in the body of the disk.  Inside the
marginally stable orbit, this ratio increases slowly in magnitude as
radial accretion ``combs'' the field out into radial loops that are
strongly sheared, increasing the correlation between $B_R$ and
$B_\phi$.  In fact, the Maxwell stress is so correlated with the
average magnetic pressure $\langle B^2/8\pi\rangle$, that a figure
showing the stress is almost indistinguishable from one showing the
magnetic pressure.

\subsection{Accretion rate}

Perhaps the single most basic quantity of interest is the accretion
rate, $\langle \dot M\rangle$, defined in equation (\ref{mdot}).
The disk is never perfectly time-steady; $\rho v_R$ varies
substantially from place to place and from time to time.  In fact, disk
quantities tend to be quite nonaxisymmetric, and much of the final
accretion inside $r_{ms}$ takes place in a spiral flow.  However,
time-averaging can remove much of this variability, and
Figure 5 shows the time-averaged accretion rate as a function of $R$
along with the instantaneous values of $\dot M (R)$ at $t= 1000$
and 1500.  The time-averaged $\dot M(R)$ is nearly constant as a
function of radius inside $R=10$, with $\dot M = 5$.  We shall adopt
this as the fiducial steady state accretion rate for the inner disk.
Between $R=9$ and 14 there is inflow at a decreasing rate; outside
$R=14$ the net flux is outward.  In this simulation, the initial disk
is entirely contained within the grid, and the outer part of the disk
must move outward in order to ``soak up" the angular momentum received
from the inner disk material.  Over the course of the simulation the
total disk mass decreases by 14\%; about 90\% of this has gone into the
black hole.

Figure~6 shows a detailed time-history of the accretion rate through
the inner radial boundary.  As can be seen from this figure, the
accretion rate shows significant time variations after its initial
growth phase.  The nature of the time-variability can be studied using
the Fourier power spectrum of the accretion rate
during the latter portion of the simulation (beyond $t=600$, i.e., after the
initial growth phase is over).  This is shown in Figure~7.  It is
most simply described as a ``red" power-law; roughly speaking,
$P(f) \propto f^{-1.7}$.  However, there is also some curvature, in the
sense that it steepens with increasing frequency.  Because the accretion
rate does appear to have a well-defined mean value, $d\ln P/d\ln f$ must
be $> -1$ at the lowest frequencies.  At the high frequency end,
the spectrum is largely featureless, extending almost to the inverse
free-fall time at the innermost boundary.  Other than a smooth
steepening, there is no special feature at the orbital frequency
of the marginally stable orbit (marked on the figure).  The flow is
unsteady everywhere, even well inside the plunging region.

Figure~6 provides some idea of how nonsteady the accretion is; the mass
flux is highly spatially inhomogeneous as well.  To gain some sense of
the magnitude of the spatial variations, we show the accretion
rate as a function of position on a cylinder at $R= 5$ (Figure~8).  As
this figure shows, the fluctuations are very large.  If we suppose that
the mean accretion rate of 5 is evenly distributed over the range of
altitudes $-1 \leq z \leq 1$, the mean mass flux would be $\sim 0.1$.
As the figure makes plain, there are several places where the mass flux
is five times as great as this; there are also several places where the
mass flux is $\simeq 0.3$ -- 0.4 {\it outward}.  Relatively large
fluctuations compared to the mean are to be expected when the transport of
angular momentum in the disk is by turbulence. 

\subsection{Propagation of fluctuations}

  As we have stressed all along, one of the most striking features
of this simulation is the presence of large amplitude nonaxisymmetric
modulations in the density, mass flux, and magnetic field strength.
We have already discussed how $(\phi,t)$ spacetime plots enable us to
measure their angular velocity.  It is possible to learn more
by constructing analogous plots in $(R,t)$.  For example, in Figure~9
we show the history of the azimuthally- and vertically-averaged
mass flux (eq. \ref{mdot}) in this simulation.  The diagonal stripes
show the radial motion through the disk of fluctuations in the mass
flux, their slopes indicating their radial group speeds.  Typically
these disturbances begin near $R \simeq 5$ and travel both outward and
inward.  In the outward direction, they move at a very nearly constant
speed $\simeq 0.07$, while in the inward direction, their speed is at
least three times faster.  Because the outward speed very nearly
matches the magnetosonic speed, we infer that these disturbances
propagate as magnetosonic waves in the outward direction.  On the other
hand, inward-going mass flux perturbations are simply advected along
with the dynamical inflow.

   We conjecture that their origin lies in the way in which material
enters the region of unstable orbits.  When a fluid element begins to
accelerate inward, it creates a rarefaction wave that moves outward at
the magnetosonic speed, while traveling azimuthally at the (faster)
orbital speed.  On the other hand, the radial inflow speed
near and inside the marginally stable orbit is comparable to the
magnetosonic speed, so fluctuations travel inward rather more quickly.
These fluctuations are never smoothed out into a completely steady flow
because individual fluid elements move into the plunging region as a
result of the continually changing torques exerted by the MHD
turbulence (see the next subsection).  Thus, at this level of detail,
there is no way to achieve a stationary state.

    Although strong nonaxisymmetric pressure waves were generated by MHD
turbulence in the shearing box simulations, they had no
significant impact on the evolution of those simulations.  In this
global simulation, however, as Figure~9 shows, they play
a significant role in the creation of large fluctuations in the
accretion rate.

\subsection{$R$--$\phi$ Stress}

The total stress must satisfy the equation of angular momentum
conservation.  In a steady-state disk, this means that locally the
torque must be equal to the mass accretion rate (which is constant)
times the difference between the local specific angular momentum $\ell$
and the specific angular momentum carried into the black hole
$\ell(r_{g})$.  Thus, the stress is equal to
\begin{equation}\label{steadystress} 
\bar
S = {\dot M \Omega(R) \over 2\pi} \left[1 - {\ell(r_g) 
\over \ell(R)}\right], 
\end{equation} 
where $\dot M$ is the total mass accretion rate, $\Omega(R)$ the
rotational frequency (for the pseudo-Newtonian potential, $\Omega(R) =
1/[R^{1/2}(r-1)]$), and $\ell$ is the specific angular momentum ($\ell
\equiv R^2 \Omega$).
In the time-steady, axisymmetric ``Novikov-Thorne" disk, it is assumed
that $\ell(r_g) = \ell(r_{ms})$ because the stress is assumed to be zero
for all $R \leq r_{ms}$.

For the purpose of studying angular momentum transport, it is useful to
look at the azimuthally- and vertically-integrated magnetic stress,
$\langle \int dz  M^{R\phi}\rangle_\phi$, as this is the quantity most directly
related to transfer of the $z$-component of angular momentum.  In particular,
we compare this quantity to the total stress per unit area
that would be expected in the steady state ``Novikov-Thorne" disk.  As
discussed above, in the inner part of the disk the mean accretion rate
is about 5 in code units.  In Figure~10 $\bar S$ (for $\dot M=5$ and
$\ell(r_g) = \ell(r_{ms})$) is
contrasted with $\langle \int \, dz \, M^{R\phi}\rangle_\phi$ averaged
from $t=750$ to $t=1500$.  As that figure
shows, in the main body of the disk (approximately $4 \leq R \leq 15$),
the average magnetic stress is almost exactly equal to the expected stress.
At large radius the two curves diverge because
equation (\ref{steadystress}) assumes a time-steady disk, whereas ours
has only a finite mass, so the behavior at its radial exterior is quite
different.  At the inner edge of the disk, in contrast to the
prediction of the conventional model, the azimuthally- and vertically-averaged
magnetic stress does not
diminish, but, if anything, increases slightly inward (the final upward
spike is an artifact of the boundary condition).  
Note, however, that the local value of the ratio
$|M^{R\phi}/(\rho v_R v_\phi)|$ in the plunging region varies widely:
from as much as $\sim 10$ several scale-heights out of the equatorial
plane to as little as $\sim 10^{-4}$ at some locations in the plane.

   In the context of time-steady and axisymmetric disks, the stress is often
discussed in terms of the dimensionless parameter introduced by Shakura
\& Sunyaev (1973),
\begin{equation}
\alpha_{SS} \equiv \langle {\int \, dz \, T^{R\phi} \over \int \, dz \, p}
                   \rangle_\phi .
\end{equation}
In many papers following this seminal work, the ``$\alpha$ model" is
generalized so that the {\it local} ratio of stress to pressure
is measured by the quantity $\alpha \equiv |T^{R\phi}|/p$.  Often,
it is also assumed that the stress is due to some variety of
turbulent viscosity.  However,
here we follow the spirit of the original Shakura \& Sunyaev work
by supposing that $\alpha$ is merely a convenient dimensionless
way of measuring the stress in units of the pressure.  We do not
particularly care whether the stress is due to chaotic or regular
mechanisms.  In this spirit,
we define $\alpha \equiv |M^{R\phi}|/p$.
Unlike $\alpha_{SS}$, which, in
most studies, is taken to be a constant independent of time and location,
$\alpha$ in this simulation is a dynamical and highly variable quantity
that is determined by local and transient magnetic dynamics.  For this
reason, the absolute value is significant; although $M^{R\phi}$ is
positive most places, there are counter-examples.

To give a sense of its scale and range of variation, we first examine
$\langle \alpha \rangle_\phi$, the azimuthally-averaged version of this
ratio, as found in the late-time snapshot (Figure~11).  In the main
portion of the disk, $\langle \alpha \rangle_\phi$ ranges between $\sim
10^{-2}$ and $\sim 10^{-1}$.  However, it rises dramatically (to
$\sim10$) at several scale-heights above and below the mid-plane.  The
same tendency has been seen (with much better vertical resolution)
in stratified shearing box simulations (Miller \& Stone 2000).
$\langle \alpha \rangle_\phi$ also increases sharply at small radius.
In the mid-plane, $\langle \alpha \rangle_\phi \simeq 0.2$ at $R=3$,
but rises to $\sim 3$ at the inner radial edge of the simulation.
Moreover, the vertical
rise in $\langle\alpha\rangle_\phi$ also becomes sharper at small
radius, so that at $R=3$, $\langle \alpha \rangle_\phi \sim 10$ at
$|z|$ as small as 0.5.

In addition to these systematic trends, $\alpha$ is also subject to
sizable fluctuations.  Measured in the same way as for the magnetic
energy density and the gas density (i.e., in terms of fluctuations in
azimuth at fixed $R$ and $z$), the {\it rms} fractional fluctuation
amplitude in the bulk of the disk ranges from $\simeq 1.5$ to $\simeq
3$, but at large altitude above the disk midplane, the fluctuations can
be much greater.  The character of these fluctuations is illustrated by
a different projection of $\alpha$, which we call $\bar{\alpha}$:
\begin{equation}
\bar{\alpha} \equiv {\int \, dz \, \rho \alpha \over \int \, dz \, \rho},
\end{equation}
the density-weighted vertical average of $\alpha$.  In this projection
(Figure~12), we clearly see the nature of the azimuthal fluctuations:
just as for the density fluctuations (shown on smaller scale in Figure~2),
the magnetic fluctuations are sheared filaments of limited angular coherence
length.

These results highlight the limitations of the primary assumption made
in many previous studies of disk structure, that $\alpha_{SS}$ is a
universal constant, independent of time and location.  By contrast, we
find that the ratio of stress to pressure varies strongly from place to
place, both in the sense of having large fluctuations and in the sense
of possessing strong systematic gradients.  In many places the gas
pressure is relatively smooth while the magnetic stress varies
considerably.  In other places the variation in $\alpha$ is simply due
to changes in the background pressure.  For example, the magnetic
stress does not drop with height as rapidly as the pressure does.

For purposes of comparison with standard disk models, it is possible to
compute (albeit with some trepidation) the value of $\alpha_{SS}$ found
in this simulation.  This quantity, defined in terms of the Maxwell
stress, is shown as a function of radius in Figure~13 for the disk at
the end of the simulation.  It is $\sim 0.1$ through most of the disk,
but, starting just outside the marginally stable orbit, it rises inward.
It reaches a peak $\sim 1$ near the inner radial boundary of the
simulation.   The rise is primarily due to the rapid drop in gas
pressure as the accretion flow accelerates inward.  Also shown is the
ratio of the Maxwell stress to the magnetic pressure (dashed line).
This $\alpha_{mag}$ is 0.2--0.3 throughout most of the disk, and
increases inside $r_{ms}$.

Whether $\alpha_{SS}$ is meaningful depends on how it is used.  As the
previous discussion emphasized, the local ratio of magnetic stress to
pressure has strong vertical gradients that are completely obscured
when the vertically-integrated version is used.  Similarly, there are
large fluctuations in azimuth and as a function of time.   For these reasons,
$\alpha_{SS}$ is useless as a local dynamical variable.  However,
the approximate constancy of $\alpha_{SS}$ as a function of radius {\it
in the disk proper} means that it can be used as a rough estimator of
the surface density in the body of the disk, provided its use is not
extended to local properties nor is it thought to be accurate at better
than ``factor of a few" accuracy.  This conclusion is entirely
consistent with many standard uses of $\alpha$, and within the spirit
with which it was originally introduced by Shakura and Sunyaev (1973)
as a scaling parameter, not a fundamental variable.

\subsection{Accreted angular momentum and energy}

Another significant diagnostic of the flow is the mean specific angular
momentum distribution.  There are a variety of ways to characterize the
specific angular momentum; one is by vertical and
azimuthal average, $\langle \ell \rangle$, as defined in equation
(\ref{el}).  This function was discussed in Hawley (2000) for the model
GT4 (see Fig.~14 in Hawley 2000);  the present model is very similar.
At late time the average specific angular momentum is slightly
sub-Keplerian throughout most of the disk.  The most striking feature
is that $\langle \ell \rangle$ continues to decline inside $r_{ms}$
rather than leveling off.  If all stress ceased at the marginally
stable orbit, material would travel inward from there retaining
precisely its energy and angular momentum.  Conversely, the degree to
which fluid changes its specific energy and angular momentum is a
measure of the stresses that exist in the plunging region.  The mean
specific angular momentum diminishes from 2.60 at $r_{ms}$ to 2.48 at
the innermost radius.  Thus, on average in this snapshot, the accreted
matter has lost $\sim$5\% of the angular momentum it had when it
was at the marginally stable orbit.  It is likely (for reasons
explained in \S 5) that this number is an underestimate, and in reality
the specific angular momentum at the innermost radius is rather
smaller.

The total rate at which angular momentum is carried across the inner
radial boundary is slightly smaller than these numbers would suggest
because there is also a small outward electromagnetic angular momentum
flux.  This electromagnetic flux is due to a generic property of
magnetic fields in shearing plasma: when the orbital frequency
decreases outward, the field is swept back so as to transport angular
momentum outward, i.e., $\langle B_R B_\phi \rangle < 0$.  However,
near the inner radial boundary this is a small effect, the
electromagnetic angular momentum flux there is only about 2\% of the
angular momentum flux carried by the matter.

Further complexity is revealed by a closer examination of the $(R,z)$
distribution of the late time azimuthally-averaged angular momentum
inside the plunging region.  In fact, the specific angular momentum
in the cells lying along the equatorial plane
is nearly constant inside $r_{ms}$.  However, the value of
$\ell$ drops with $z$ above and below the equator.  This is consistent
with the distribution of magnetic field which, of course, accounts for
the nonzero stress.  The low value of field near the equator is in part
a consequence of the assumed symmetry of the initial distribution of
field.  The initial field loops are symmetric with respect to the
equator ($B_R$ and $B_\phi$ are zero there), and throughout the run the
strongest fields tend to be located above and below the equatorial
plane.

Just as for every other quantity, there is also considerable spiral structure.
Figure~14 shows the density-weighted vertically-averaged specific
angular momentum.  It makes plain how non\-axisymmetric the distribution of
torque is in the plunging region.

    For a number of reasons, the accreted energy is harder to determine
than the accreted angular momentum (see further discussion in \S 5.2).
To introduce this topic, we begin by displaying (Figure~15) the
specific energy $e$ in rest-mass units,
\begin{equation}\label{specenergy}
ec^2 = {1\over 2} v^2 + {(\Gamma -1)P\over \rho } 
+ {B^2\over8\pi \rho } - {1\over (r-r_g)},
\end{equation}
in the equatorial plane at late time.  The specific energy for a
circular orbit precisely at $r_{ms}$ is -0.0625; here, $\langle e
\rangle_\phi$ is a bit greater than that ($\simeq -0.055$) because of
the thermal, magnetic, and turbulent kinetic energy contributions to
the total energy.  In the equatorial plane, $\langle e \rangle_\phi$ is
almost constant from $R=3$ to just outside $R = 2$, but actually {\it
rises} as $R$ decreases from there to $R=1.6$, reaching a peak that is
very nearly unbound: -0.01.  The azimuthally-averaged energy in the
equatorial plane then falls sharply to -0.07.  Even this wild swing
underestimates the fluctuations: the smallest specific energy found in
the equatorial plane is -0.19.  Thus, we are confident in asserting
that there is a large amount of energy transfer within the plunging
region.  However, given these large fluctuations, we do not feel that
we can compute a well-defined mean energy for the accreted matter.  It
is likely, moreover, that at least some of the very large fluctuations
in the innermost radial zones are not physical, but due to the inner
boundary condition.

In fact, the situation is somewhat worse than this.  The reason is that
dissipation is crucial for computing the accreted energy, and our
simulation contains no real dissipation physics.  Absent numerical
losses (see \S 5.1, 5.2), energy taken from plunging matter may go
several places.  One possibility is to build magnetic field energy in
the plunging region itself.  If the field energy is advected into the
black hole without dissipation and radiative losses, the transfer of
energy from fluid to magnetic field makes no difference to the accreted
energy.  It is also possible for magnetic stresses to transfer energy
all the way back to the disk proper (and indeed we see significant
magnetic stress throughout the plunging region).   Again, without
dissipation, the energy transferred in that way stays in the fluid,
whether in the form of kinetic or magnetic energy.  Only in the event
that the energy is transferred all the way out to non-accreting matter
would it be possible for that energy to avoid being carried back in.
Thus, to the degree that this simulation avoids numerical losses, it is
very difficult for it to show any significant transfer of energy out of
the plunging region.  By contrast, in a real disk, dissipation changes
energy into heat, which may then be radiated if the thermal timescale
is shorter than the inflow timescale; in our simulation, neither
physical dissipation nor radiation is present.  

\section{Discussion}

\subsection{Black hole spin-up}

We are now ready to begin answering some of the questions posed in the
Introduction.  Most significantly, this simulation has shown that the
stress due to MHD turbulence does not automatically cease at the
marginally stable orbit.  In fact, it continues to be present
throughout the plunging region.  As an immediate consequence, matter
continues to lose angular momentum as it travels from the marginally
stable orbit toward the event horizon.  Because it then enters the
black hole with smaller angular momentum than if no stresses had been
present, the rate at which the black hole is spun up is slower than it
would have been otherwise (Agol \& Krolik 2000; Gammie 1999).  It
further follows that estimates of the black hole spin distribution are
shifted toward slower rotation (e.g., Moderski \& Sikora 1996).
Although in this simulation the magnitude of this effect is small
($\simeq 5\%$), it is possible that more realistic conditions will
increase its size (see \S 5.1).

\subsection{Additional dissipation in the inner disk}

    We have already remarked on the fact that our simulation does not
directly compute the dissipation rate.  In this section we attempt to
estimate what dissipation might be associated with dynamics of the sort
seen here.  In conventional disk models, the local dissipation rate is
directly tied to the local stress:
\begin{equation}\label{dissipation}
Q = \left|{d\Omega \over d\ln r}\right| \int \, dz T^{R\phi},
\end{equation}
where $Q$ is the dissipation rate per unit area.  This relationship is
predicated upon the notion that the same stress producing local
transport also produces local dissipation.  This remains true for the
mean flow dynamics of MHD turbulence (Balbus \& Papaloizou 1999); in
fact, numerical dissipation mimicking the physical dissipation at the
short wavelength end of the turbulent cascade is the way in which
matter in this simulation loses energy as it moves slowly inward in the
body of the disk.  However, this identification is not necessarily
correct as a prediction for the instantaneous rate of dissipation in
specific fluid elements.  It is especially questionable in the plunging
region.  The stress is not necessarily turbulent, and there may not be
enough time for the energy deposited in the fluid by the stress to be
transferred to small enough lengthscales for it to be dissipated.
Nonetheless, in this subsection we adopt the speculative assumption that
the vertically-integrated dissipation rate is proportional to the
vertically-integrated stress in order to speculate about the
consequences for disk radiation that may result.

One possible consequence is an altered radial distribution of the disk
surface brightness.  As shown in Figure~10, the azimuthally-averaged
stress in the simulation is nearly constant inside $R\simeq 5$.  If the
relation of equation (\ref{dissipation}) holds, the total dissipation
rate outside $R=3$ in this simulation would be $\simeq 50\%$ greater
than in a disk with the same accretion rate, but zero stress at the
marginally stable orbit.

If this were to happen, there would be a number of effects of
considerable phenomenological interest (Agol \& Krolik 2000).  Because
this additional heat is released in a relatively small area of the
disk, it leads to relatively high temperatures, and therefore
contributes to the highest-frequency portion of the spectrum.  Another
consequence stems from the fact that the dissipation is located in the
innermost part of the disk, where relativistic effects are strongest.
The light that is radiated due to this extra heating should be
comparatively strongly beamed into the equatorial plane by a
combination of special relativistic Doppler beaming driven by the
orbital motion and general relativistic trajectory-bending.
Trajectory-bending also forces a large fraction of the
photons to return to the disk at larger radius, to be reprocessed there
into lower-frequency light.

Whether or not the enhanced stresses in the plunging region 
lead to additional radiative losses
depends on the dissipation and photon diffusion rates.  In the disk
proper, dissipation takes place in the turbulent cascade, in small
lengthscale fluctuations that are short-lived, exhibit little spatial
coherence, but are smoothly distributed in a statistical sense.  In the
time-varying accretion of the plunging region this turbulent cascade
may not have time to go to completion.  On the other hand, there might
be dissipative events, mediated by reconnection for example, that
involve structures with relatively large spatial coherence lengths and
which are not necessarily smoothly distributed throughout the region.
As we will discuss in \S 5.1, our limited spatial resolution makes it
difficult for us to estimate fairly the rate of such events in the
plunging region.  Moreover, assessing the radiative properties of such
dissipative events is beyond the the capabilities of this simulation.
However, if events of this sort do occur, significant luminosity might
result because the reduced optical depth of the plunging matter (as
compared to the disk proper) also greatly reduces the thermal time.
Note that this mechanism results in an enhanced radiative efficiency
without the transfer of any energy from the plunging region to the disk
outside the marginally stable orbit.

Finally, although the azimuthally-averaged stress matches conventional
expectations in much of the disk, there are large azimuthal variations
in the form of sheared filaments (Figure~12).  Translated into a
dissipation rate, these spiral waves of enhanced (or diminished) stress
become spirals of enhanced (or diminished) heating.  Whether this
concentration of heating has an effect on the emitted radiation depends
on the characteristic timescale of these fluctuations {\it as seen in a
specific fluid element}.  If a particular fluid element has an elevated
heating rate for as long as its thermal timescale, the effective
radiating area of the disk is reduced to the area of the regions where
the heating is greatest, and the typical effective temperature is
raised accordingly.  On the other hand, if over a thermal timescale
individual fluid elements see both positive and negative fluctuations,
the effects will average out.   We have already estimated that the
characteristic speed of fluctuations through the disk is of order the
magnetosonic speed; on the other hand, the thermal timescale is $\sim
\alpha_{SS}^{-1}$ times the orbital period.  Consequently, the ratio of
the coherence time of fluctuations to the thermal time should be, very
roughly, $\sim \alpha_{SS} v_{orb}/(v_{A}+c_s)$; in this simulation this
ratio is $\sim 1$.

\subsection{Fluctuations in light output}

    We have emphasized throughout this paper the highly non-stationary
behavior that is intrinsic to disk behavior as found in this simulation.
One form these fluctuations can take is the coherent waves that
are continually radiated outward through the disk from the region of
the marginally stable orbit.  Even if, on average, dynamics in the
plunging region change the mean accretion efficiency by only a small
amount, the large fluctuations we observe in the simulation should cause
substantial time-dependent variations in the light output.

Once again, because there is no direct connection between the dynamics
traced by this simulation and dissipation, we cannot directly predict the time
variation of the disk luminosity.  In fact, to do so would also require
calculating the photon diffusion rate from inside the (optically thick)
disk.  However, some indication of what may occur may be gleaned from
Figures~6 and 7.  Some of the fluctuations in the accretion rate
(particularly those at the highest frequencies) are due to dynamics in
the plunging region; these may or may not lead to dissipation (see \S
4.2), and therefore may or may not contribute to fluctuations in the
light output.  For this reason, and following the {\it ansatz} of
equation (\ref{dissipation}), in Figure~7 we also plot the Fourier
power spectrum of the volume-integrated magnetic stress.  Whether the
accretion rate or the stress is more closely related to the dissipation
rate, the actual output is the convolution of fluctuations in the
heating rate with the probability distribution for the photon escape
time.  In frequency space, this amounts to a (position-dependent)
low-pass filter.  When the heating occurs deep inside the disk, the
cut-off frequency for this filter is $\sim \alpha_{SS}\Omega$;
fluctuations in the light due to heating closer to the surface are
subject to a less stringent frequency cut-off (in the optically thin
limit, there is no cut-off at all).  Consequently, we expect the
Fourier power spectrum of the disk luminosity to be rather strongly cut
off at frequencies above $\sim 10^{-2}$, but the power spectrum of
fluctuations associated with coronal dynamics may be different.

The observation of QPOs in black hole candidates has
motivated a number of studies of possible sources of systematic time
variability in accretion disks.  In a series of papers (Nowak \&
Wagoner 1991, 1992; Perez et al. 1997) Wagoner and collaborators have
explored the linear theory of fluid oscillations in the relativistic
portions of stationary accretion disks.  In these disks only
hydrodynamic effects are considered, and angular momentum transport is
modeled by a pseudo-viscosity parameterized by a constant
$\alpha$.  These disks possess a variety of normal modes, including
$p$-modes that are trapped between $r_{ms}$ and the radius where the
epicyclic frequency is maximum, and $g$-modes, whose greatest amplitude
is found under the peak of the epicyclic frequency curve.  In the
Paczy\'nski-Wiita potential we use, the position of the maximum
epicyclic frequency is $R \simeq 3.6$; with the additional correction
terms introduced by Nowak and Wagoner (1991), it moves outward to
$R=4$.  We would not expect to find exactly the same mode frequencies
as they because of the difference in the potentials used; however, the
mode frequencies should in general be $\sim \Omega(r_{ms})$.  A perusal
of Figure~7 fails to find any evidence for special frequencies in this
range.

There are several reasons why trapped disk oscillations do not appear
here.  First, the normal modes discussed in the previous paragraphs are
predicted on the basis of a purely {\it hydrodynamic} analysis; no
account is made of magnetic forces.  Here, however, magnetic forces
play a major role in determining the dynamics of fluid elements.  For
example, if one repeated the derivation of the ``diskoseismic"
dispersion relation, but allowed for a weak magnetic field, one would
rediscover the magneto-rotational instability, thus vitiating the
hydrodynamic ``normal modes".  More fundamentally, our simulation
indicates that there is no underlying quiet flow to support coherent
oscillations.  The disk is so turbulent that there is no steady
equilibrium against which linear perturbations can develop.  Put
another way, we find that the typical radial motion timescale in the
vicinity of $r_{ms}$ (where these normal modes are supposed to be
concentrated) is comparable to the predicted mode frequencies, so the
fundamental assumption that these modes are perturbations to circular
orbits is inappropriate.

\section{Limitations}

The results presented here and in Hawley (2000) obviously represent
only the first steps in detailed simulations of global accretion
disks.  In this section we outline how the approximations we have made
may impact our conclusions.  Of course the only definitive way to
investigate the effects of these approximations would be with
additional simulations, and this discussion should be viewed as a guide
for future work.

\subsection{Effects of resolution}

Foremost among the limitations is the numerical resolution of the
simulation.  Although we placed 30 radial zones between $R=1.5$ and
$R=4$, that is probably insufficient.  There are two reasons for
believing this to be the case.  First, we may apply the standard test
for convergence of comparing two simulations with the same physics but
different resolution, and ask whether they show significant
differences.  Simulation GT4, reported in Hawley (2000), and a new
$64^3$ run (hereafter referred to as run LR for Lower Resolution) serve
as good comparisons.  Both GT4 and LR share identical initial and
boundary conditions.  GT4 had equally spaced grid zones over a distance
of 20 in both $R$ and $z$ for $\Delta R = \Delta z = 0.156$.
Simulation LR used equally spaced radial zones over a range of 30 for
$\Delta R = 0.47$, and used half of the $z$ grid zones centered around
the equator within $z = \pm 2$, for $\Delta z= 0.125$.  By contrast,
here $\Delta R = 0.0833$ and $\Delta z = 0.0625$ in the inner region.

All three simulations are qualitatively similar in global properties,
as measured by volume integration, but differ in certain other respects
that are important to physical interpretation.  The time-averaged
accretion rate in GT4 is about 40\% smaller than in the work reported
here (although the time-averaged accretion rate in LR is only about 2\%
smaller).  However, the most striking contrast is in magnetic field
strength.  Higher resolution results in larger magnetic fields,
particularly in the poloidal components.  Contrasting maps of the
magnetic field intensity at similar late time for GT4 and our new
simulation (Figure~16), we find that the field is consistently stronger
in the new simulation everywhere inside $R=5$, rising to typically a
factor of 3 greater in magnitude inside $R=4$.  Significantly, outside
$R=5$, the field is generally stronger in the old simulation (sometimes
by more than a factor of 10), but that, too, is in keeping with the
sense of the resolution contrast: GT4 has better resolution in the
outer regions.

The field intensity is controlled by a trade-off between the turbulent
MHD dynamo driven by the magneto-rotational instability and numerical
dissipation (recall that there is no explicit resistivity in this
simulation).  Whatever the real physics of resistivity in accretion
disks may be, it is unlikely that it is as large as the effective
numerical resistivity induced by our modest spatial resolution.  The
contrast between the field intensity in the two simulations is
therefore very likely explained by the smaller numerical resistivity of
the better-resolved simulation.

Another piece of evidence supporting the belief that the magnetic field
is weaker than it would be with better resolution is presented in
Figure~17, where we display the magnetic field structure in the
equatorial plane within $R=4$, plotted for every fourth grid zone.  
The planar component of the field is
organized into long sheared loops.  Because the loops are wrapped
around one another, there are many places where the field direction
reverses within 4 -- 8 grid cells.  This means that with a modest
amount of flow convergence numerical effects could cause the field to
decay on a relatively short timescale.

Finally there is the issue of the computational domain employed, in
particular the location of the boundaries.  Very little outflow
leaves through the $z$ boundaries (typically only a few
percent of the accretion rate).  The outer radial boundary was
extended in the present simulation compared to GT4, which had the
effect of keeping more of the disk on the grid, but otherwise had no
obvious effect on the evolution.  We expect that there is also little
dependence upon the location of the inner boundary because the
flow across that boundary is generally supermagnetosonic.
Although we made no tests here of the effect of varying $r_{in}$, tests
of this sort were made on the similar simulations in Hawley (2000),
and it appeared to make little difference.

\subsection{Energy conservation and physical dissipation rates}

The artificial dissipation caused by insufficient resolution also has
consequences for energy conservation.  To very good accuracy, angular
momentum is conserved in this simulation.  The primary source of
numerical dissipation, the truncation error associated with transport
terms in the equation, does not alter the total angular momentum,
although it may artificially change its distribution.
Total energy, however, is different because the energy equation that is
solved is the internal energy equation.  Numerical dissipation can
destroy kinetic energy in regions of strong velocity gradients and
magnetic energy in regions of strong magnetic field gradients, but
there is no provision in this simulation for putting that energy back
into heat (except for dissipation in shocks through the use of an
artificial viscosity).  Both sorts of dissipation can also occur in
real systems, but at rates that might be very different from what
happens in the simulation.

In the body of the disk, the problems that numerical dissipation can
cause are limited.  The single largest items in the energy budget are
potential energy and rotational kinetic energy; neither one is
significantly affected by numerical effects.  The thermal energy is
only a small fraction of those energies; this is a prime motivation
for solving an internal rather than total energy equation.
Artificial energy losses occur in the dissipation of relatively small
amounts of energy bound up in small-scale random fluid motions and in
magnetic field.  In fact, in this regard, numerical dissipation acts in
a way that is consistent with the physical dissipation of energy at the short
lengthscale end of the turbulent cascade.  In both cases, the losses
occur on the shortest lengthscales, and in both cases (assuming the
disk radiates efficiently) energy is lost
from the disk on timescales short compared to the inflow time.

However, near and inside the marginally stable orbit, detailed
questions of energy transfer become significant.  The contributions to
the energy budget due to both turbulent motions and magnetic field
grow.  At the same time, their importance also grows.   For example, it
would be of great interest to determine whether plunging fluid elements
can, by exerting magnetic stresses, transfer energy back to the disk
proper (cf. \S 3.7).  As we have seen, significant magnetic stresses
exist throughout this region, so there is some possibility that this happens. 
However, the rate at which this may occur is regulated by the magnetic field
strength and topology, and these are subject to significant artificial
numerical dissipation.  In addition, any energy that is delivered is
likely to take the form of MHD turbulence; whether this turbulence is
dissipated (and the energy radiated) quickly or slowly compared to the
inflow is critical for evaluating the radiative efficiency that
results.  Because the inflow, thermal, and dissipation timescales all
change significantly in the vicinity of $r_{ms}$, it is unlikely that
in this region numerical dissipation serves as a successful stand-in
for physical dissipation and radiative losses in the way it does at
larger radii. 

\subsection{Equation of state}

Dissipation and radiation are, of course, key elements in the gas
thermodynamics.  In the present simulation, we describe the gas's
internal energy in only an approximate way: we assume it is
adiabatic except for heating due to artificial viscosity.
Explicit treatment of dissipation and radiation would fill in the
missing terms in the gas energy equation.

In the absence of a full treatment of the energy equation, we can
at best make educated speculations about the impact of our assumption
that the gas behaves adiabatically. Some consequences were already
discussed in the previous subsection.  There are several more, not
as closely related to resolution issues.

As we have remarked earlier (\S 3.7), one of the most dramatic features
of this simulation is the coherent disturbances that propagate as
magnetosonic waves from the vicinity of the marginally stable orbit
outward through the disk.  Their behavior depends strongly, of course,
on the sound speed in the disk.  However, in this simulation, in which
all gas is given the same specific entropy initially and then evolved
adiabatically, the sound speed could be quite different from what it
would be in a more realistic simulation that includes heating from
dissipation of MHD turbulence and cooling via radiation.  We
expect, therefore, that properties that depend on pressure forces (such
as the propagation of these waves, or the vertical thickness of the
disk) are not well-described in this simulation.

Another consequence of our assumed equation of state, combined with
the particular value of specific entropy we assigned the gas, is that
the disk is modestly thick.  As a result, it begins from a non-Keplerian,
partially pressure-supported state and radial pressure gradients
drive some of its initial evolution.   Magnetic stresses
rapidly change the angular momentum distribution within the torus to
nearly Keplerian, but without cooling, the internal pressure continues to
push matter away from the pressure maximum.  Some accretion,
therefore, is driven not by magnetic stresses but simply by the pressure
gradient.  At late time, however, the inner disk establishes a
quasi-steady state and nearly all the accretion can be accounted for by
the Maxwell stress.  It is likely that the observed dynamics are
reasonably representative of generic inner disk accretion.  

Finally, although the present disk is moderately thin, with $H/R
\simeq 0.15$ in the inner region, it is conceivable that new
phenomena might appear in a still thinner disk.  Although our results
provide no particular reason to predict this, it is possible that in a
disk with still smaller $H/R$ there might be modes with wavelength
large compared to $H$ yet still small compared to $R$.  Questions such
as these could be investigated by contrasting the present work with
simulations of the inner regions of more extended, thinner, initially
Keplerian disks.

A possible example of this sort of comparison is provided by the work
of Armitage et al. (2000).  Like us, they also simulated the inner
region of an accretion disk in a pseudo-Newtonian potential using a 
three dimensional
MHD code.  They find a smaller reduction of $\ell$ inside $r_{ms}$ than
we find in the simulation reported here, an effect they conjecture
could be due to the smaller temperature and scale height in their
disk.  However, their scale height is $\simeq 0.08$, not much smaller
than our typical value of $\simeq 0.15$.  Moreover, there are other
differences between the two simulations that may also account for
quantitative contrasts: in their stratified simulation, only a limited
vertical extent is simulated; their equation of state is isothermal;
and they simulate a $\pi/3$ wedge, rather than a full $2\pi$ in
azimuth.  Because in our simulation fluctuations with azimuthal
wavelength $\sim 1$~radian or more have considerable amplitude, the
limitation to a wedge could have significant consequences.  In any
event, as they emphasize, the differences between their simulations and
ours are simply quantitative, not qualitative.  They, too, find a
nonzero Maxwell stress inside $r_{ms}$, but its amplitude is lower.
Further work is required to determine what sets the stress levels in a
variety of different situations.

Thus, for all these reasons, it will be important for future simulations
to incorporate improved treatment of gas heating and cooling.  It
should not be difficult to integrate a total energy equation along with
the gas momentum (i.e., force) equation.  Heating could be treated in a
controlled way by inserting an explicit resistivity into the magnetic
field evolution equation and an explicit viscosity into the fluid force
equation, and placing corresponding terms in the internal energy equation.
Radiative losses could be described with either an escape probability
formalism, or, in the longer term, solution of the radiation transfer
problem.

A simulation with both  heating and radiation would simulate
the behavior of thin, radiative disks.  There has been no previous
work of this sort; the significance of such an effort is obvious.

Heating alone would, in effect, lead to a simulation of the popular
advection dominated accretion flow (ADAF) model.  Some previous
simulations of this model have used a phenomenological prescription for
viscosity (e.g., Igumenshchev \& Abramowicz 1999; Igumenshchev,
Abramowicz, \& Narayan
2000; Stone, Pringle \& Begelman 1999).  Others have computed the
stress from a global magnetic field (in axisymmetry: Uchida \& Shibata
1985 and Stone \& Norman 1994; in three dimensions:  Matsumoto 1999).
Still another approach has been to study the stress created by MHD
turbulence including resistive dissipation, but in axisymmetry (Stone
\& Pringle 2000).  Addition of explicit dissipation to the simulation
presented here would, in effect, extend the Stone \& Pringle work to
three dimensions.

\subsection{Duration} 

The timestep of the simulation is limited by the Courant condition.
The most stringent limitation comes from the high orbital velocities
and the radial plunging speeds in the inner part of the grid where the
resolution is reasonably fine.   The total duration of the simulation
was 1500 time units, requiring almost 400,000 timesteps.  As discussed
in \S3.1, this is long enough for a quasi-steady state to be
established in the inner part of the disk.  Since $\alpha \sim 0.1$,
the simulation covers several accretion times in the inner part of the
disk.  Whether or not the results obtained are truly representative,
must, however, be tested with simulations of longer duration.  An
evolution beginning with a nearly Keplerian disk that extends out to
large radius could allow the inner disk region to establish itself
self-consistently from a long-term accretion flow.

\subsection{Field topology}

What are the limitations due to the choice of initial disk and field
topology?  The initial poloidal field loops used here are amenable to
rapid development of the MRI on large scales.  Further, the presence of
significant initial radial field quickly generates through simple shear
amplification a toroidal field of near-equipartition strength.  Thus,
the initial field topology guarantees magnetically-driven evolution
within a few orbits.  In earlier work (Hawley 2000) it was shown that
in this regard (generating magnetically-driven inflow after a few orbits),
initially poloidal and initially toroidal field configurations 
are essentially
equivalent.  However, the symmetry inherent in this
choice of initial field topology is partly responsible
for the low field intensity near the equatorial plane and the weak
dipolar structure characteristic of the low-density regions (\S 3.3).

     Whether the field is contained entirely within the problem area,
or, equivalently, whether there is zero net flux through the box,
has a greater qualitative impact.  Shearing box simulations (Hawley,
Gammie \& Balbus 1995, 1996) also found strong similarities between
simulations with initial toroidal field and simulations with random
initial fields.  However, they found dramatically different behavior
when there was net vertical field.  Such field topologies tend to
produce larger net angular momentum transport and more violent
evolution.  These effects have also been observed in 
global simulations (Matsumoto 1999).

\subsection{Displacement current}

The equations of MHD as we have written them (see especially
equation~2) omit the displacement current term.  This approximation is
valid so long as the Alfv\'en speed $v_A < c$.  This condition is
satisfied almost everywhere in this simulation, with the only exception
a small region near the inner radial boundary ($1.5 \leq R \leq 2$ and
$1 \leq |z| \leq 2$).  In the equatorial plane, where most of the mass
flows, the largest value of $v_A \simeq 0.3$, and that occurs only at
the inner radial boundary; everywhere outside $R = 2$ in the midplane,
$v_A \leq 0.15$.  Future simulations in which the displacement current
term is included should be somewhat more accurate, but we doubt that
this correction will cause any major changes.

\subsection{Relativity}

Obviously, it would be preferable to run the simulation using genuine
general relativistic dynamics, rather than the approximation of
Newtonian physics in a Paczy\'nski-Wiita potential.  We might hope that
this approximation is qualitatively reasonable when attempting to mimic
dynamics around a non-rotating black hole; after all, the most
important aspect of this case is the existence of dynamically unstable
orbits inside a critical radius and enforced inward radial motion
within a still smaller distance from the center.  Further, Gammie \&
Popham (1998) compute the growth rate of the MRI in full relativity and
have shown that it remains undiminished $r_{ms}$.  The
Paczy\'nski-Wiita potential, however, has no representation of the
frame-dragging created by a rotating black hole, so it cannot be a good
approximation to accretion dynamics around a Kerr black hole.  Gammie
(1999) and Gammie \& Popham (1998) examine some of the interesting
possibilities created by the Kerr geometry, but examination of those
effects must await a code with true relativistic dynamics.

\section{Conclusions}

The simulation reported here represents another step towards more
realistic simulations of accretion disk dynamics.  By improving the
grid resolution in the inner region, we have shown that magnetic
effects become increasingly important near and inside the radius of the
marginally stable orbit.   Contrary to the most widely-held assumption,
the $R$--$\phi$ component of the stress tensor does not go to zero at
$r_{ms}$; averaged over azimuth, height, and time, it is almost
constant as a function of $R$ for $R \leq 2 r_{ms}$.  As a result,
matter in the region of unstable orbits does not follow simple energy-
and angular momentum-conserving free-fall trajectories, although we are
not yet ready to make quantitative predictions about how much the
energy and angular momentum change.

The ratio $T^{R\phi}/P \equiv \alpha$ is commonly used as a
measure of the magnitude of the stress, and almost as commonly is
thought to be a constant, independent of location and time.  Our
simulation showed clearly that while it can provide a good
qualitative measure of the stress when averaged over several orbits
in time,  at any particular instant $\alpha$ has large systematic
gradients (as a function of both $R$ and $z$) and substantial
temporal fluctuations.  On average the value of $\alpha$ in the
disk is $\sim 0.1$.

This simulation has shown that substantial time- and space
variations are the rule.  The inner regions of accretion disks around
black holes are highly turbulent, non-steady systems.  Large-amplitude
fluctuations sheared into spiral fragments come and go, many of them
initiated in the region immediately outside the marginally stable
orbit.  As a result, there are continuing fluctuations in the accretion
rate; we expect that some of these fluctuations will be mirrored in the
light curves of accreting black holes.

Future simulations, guided by these results, will probe the beyond the
limitations of the present simulation.  Within the context of the same
pseudo-Newtonian potential and the present adiabatic equation of state,
we can explore the effects of different initial disk configurations and
magnetic field strengths and topologies, at higher resolution and
for longer time.  Beyond this, it will be important to incorporate
improved physics into the models.   These include general relativistic
dynamics and improved treatments of the equation of state and
dissipative processes.   This future work should bring us still closer
to a quantitative understanding of the dynamics of accretion onto black
holes.

\begin{acknowledgments}

JHK would like to thank Eric Agol for numerous helpful conversations,
and Nancy Levenson for invaluable instruction in the wiles of IDL.
He would also like to thank Giora Shaviv for comments on the difference
between heating rate fluctuations and luminosity fluctuations.
JFH thanks Charles Gammie for useful comments on the manuscript.
This work was supported by NSF grant AST-0070979, and NASA grants
NAG5-9266 and NAG5-7500 to JFH, and NSF Grant AST-9616922 and NASA
grant NRA-99-01-ATP-031 to JHK.  Simulations were carried out on the
Cray T3E system of the San Diego Supercomputer Center of the National
Partnership for Advanced Computational Infrastructure, funded by the
NSF.

\end{acknowledgments}

\clearpage
\centerline{Figure Captions}

\figcaption{Azimuthally-averaged gas density distribution at the beginning
($t=0$: panel a) and the end ($t=1500$: panel b) of the simulation.
There are 14 evenly spaced contours.  The maximum density at $t=0$ is 34.2;
at $t=1500$, $\rho_{max}=14.3$.}

\figcaption{Density in the equatorial plane for $R \leq 10$ at $t=1500$.}

\figcaption{Azimuthally-averaged magnetic pressure, $\langle
B^2/8\pi\rangle_\phi$, plotted on a logarithmic scale from $10^{-5}$
to $10^{-2}$.  The mean value of magnetic pressure
within the region where $\rho > 0.01$ is $7.1\times 10^{-4}$.
Using the mean gas pressure this corresponds to an averge 
$\beta = 13$ in the disk.
}

\figcaption{Poloidal magnetic field structure in the slice at $\phi = \pi/2$,
overlaid on density contours.  Panel a) shows the inner, high-resolution
region; panel b) shows the disk body.  The length of each arrow
is proportional to the magnitude of the poloidal field, but the scales
are different in the two panels.}

\figcaption{Accretion rate (in code units) as a function of
radius at times $t=1000$ 
and $t=1500$ (solid lines). The dashed curve is the time-average of the
accretion rate between those two times. Note that
negative accretion rate means outward flow.}

\figcaption{Mass accretion rate $\langle \dot M \rangle$ through the
inner radial boundary as a function of time.  To capture the
time-variability accurately, the accretion rate was measured every 10
timesteps in the simulation, corresponding to an average interval of
0.0037 in time.}

\figcaption{
Fourier power density of the accretion rate through the inner boundary
(solid curve) and volume-integrated Maxwell stress (dashed curve).
Frequency is in code units; the orbital frequency (i.e., the inverse of
the orbital period $\Omega(r_{ms})/2\pi$) at the marginally stable
orbit is marked with an arrow.}

\figcaption{ Accretion rate at cylindrical radius $R=5$; inward mass flow
is given a positive sign.  Note that the 
{\it local} mass flux is almost as likely to be outward as inward.}

\figcaption{Vertically- and azimuthally-averaged mass flux as a function of
radius $R$ and time.  The diagonal stripes denote radially-propagating
waves.  Reds and yellows correspond to accretion, blues to outflow.
}

\figcaption{Azimuthally-averaged and vertically-integrated magnetic
stress averaged from $t=750$ until the end of the simulation 
(solid curve) and nominal mean stress per unit area (dashed curve).}

\figcaption{Absolute value of $\langle \alpha \rangle_\phi$, in a
logarithmic scale in the snapshot at $t=1500$.
This quantity increases with decreasing radius, and
with increasing distance above or below the disk midplane.}

\figcaption{Absolute value of the vertically-averaged $\bar{\alpha}$, plotted
on a logarithmic scale for the snapshot at $t=1500$.  Note the
large fluctuations, even when values are compared at fixed radius.
$\bar{\alpha}$ is large at both large and small $R$, but for different
reasons (see text).}

\figcaption{Shakura-Sunyaev $\alpha_{SS}$ as a function of radius 
(solid line) constructed from vertical and azimuthally  averaging the
Maxwell stress and the gas pressure.  Also shown is the averaged
$\alpha_{mag}$ as a function of radius (dashed line), constructed using
the magnetic pressure in place of the gas pressure.  $\alpha_{mag}$ is
relatively constant in the disk, whereas $\alpha_{SS}$ varies with the
gas pressure.
}

\figcaption{Density-weighted vertically-averaged specific angular momentum
in the inner portion of the accretion flow.}

\figcaption{Specific energy of matter (in rest-mass units) in the
equatorial plane within $R=4$ (the high-resolution segment of the
simulation).  Note the large amplitude spiral fluctuations.  The sharp
features immediately surrounding the inner radial boundary are 
artifacts.} 

\figcaption{Logarithm of the ratio of the density-weighted mean magnetic
field intensity in the new simulation to the same quantity in GT4 at
analogous late times.  Only the region inside $R=10$ is shown.
Because these are different simulations, one cannot expect features 
to match up precisely.  However, there is a clear general
tendency for the field in the new simulation to be stronger than in the
old simulation inside $R=5$ and weaker outside.}

\figcaption{Planar component of the magnetic field in the equatorial
slice of the high-resolution region of the simulation.  The length of
each arrow is proportional to the intensity of the magnetic field.  For
clarity, the field is shown in every fourth cell.}

\end{document}